\documentclass[a4paper,11pt]{article}
\usepackage[dvipsnames]{xcolor}
\usepackage[utf8]{inputenc}
\usepackage[bbgreekl]{mathbbol}
\usepackage{geometry}

\usepackage{xcolor}
\usepackage{amsmath, array, amssymb, amsfonts,amsthm}
\usepackage{upgreek}
\usepackage{xfrac}
\usepackage[inline]{enumitem}
\setlist{nolistsep}
\usepackage[italian, english]{babel}
\usepackage[all]{xy}
\usepackage{txfonts}  
\usepackage{sectsty}
\usepackage{booktabs}
\usepackage{caption}
\usepackage{dsfont}
\usepackage{mathtools}
\usepackage{slashed}
\usepackage[makeroom]{cancel}
\usepackage[hidelinks]{hyperref}
\usepackage{textcomp}
\usepackage{multicol}
\usepackage[title]{appendix}
\usepackage[square,numbers,sort,compress,semicolon,merge]{natbib}
\let\cite\citep 
\usepackage{hyperref}
\hypersetup{colorlinks=true, urlcolor=blue, citecolor=blue, linktoc=page}

\usepackage{tikz}
\usepackage{tikz-cd}
\usetikzlibrary{cd}
\tikzcdset{
arrow style=tikz,
diagrams={>={Straight Barb[scale=0.8]}}
}
\usetikzlibrary{matrix,arrows,decorations.pathmorphing}

\DeclareMathAlphabet{\mathpzc}{OT1}{pzc}{m}{it}

\usepackage{fancybox}

\usepackage{mathrsfs}
\usepackage{scrextend}

\makeatletter
\renewcommand*\env@matrix[1][\arraystretch]{%
  \edef\arraystretch{#1}%
  \hskip -\arraycolsep
  \let\@ifnextchar\new@ifnextchar
  \array{*\c@MaxMatrixCols c}}
\makeatother

\newcommand{\defeq}{\vcentcolon=}

\renewcommand\epsilon{\varepsilon}

\makeatletter

\newcommand{\Rmnum}[1]{\expandafter\@slowromancap\romannumeral #1@}
\makeatother

\makeatletter
\newcommand{\leqnomode}{\tagsleft@true\let\veqno\@@leqno}
\newcommand{\reqnomode}{\tagsleft@false\let\veqno\@@eqno}
\makeatother

\theoremstyle{definition}

\usepackage{braket}
\usepackage{physics}

\makeatother

\begin{document}


\title{Boson-fermion algebraic mapping in second quantization}

\author{F. \textsc{Lingua} $\,{}^{a,\,*}$ \and D. M. \textsc{Pe\~{n}afiel} $\,{}^{b,\,\dagger}$ \and L. \textsc{Ravera} $\,{}^{c,\,d,\,e,\,\star}$ \and S. \textsc{Salgado} $\,{}^{f,\,\ddagger}$ }

\date{\today}

\maketitle
\begin{center}
\vskip -0.6cm
\noindent
${}^a$ Department of Applied Physics, KTH Royal Institute of Technology -- KTH. \\
SE-10691 Stockholm, Sweden. \\[2mm]

${}^b$ Instituto de Ciencias Exactas y Naturales, Facultad de Ciencias, Univesidad Arturo Prat -- UNAP. \\
Avda. Arturo Prat 2120, Iquique, Chile. \\[2mm]

${}^c$ DISAT, Politecnico di Torino -- PoliTo. \\
Corso Duca degli Abruzzi 24, 10129 Torino, Italy. \\[2mm]

${}^d$ Istituto Nazionale di Fisica Nucleare, Section of Torino -- INFN. \\
Via P. Giuria 1, 10125 Torino, Italy. \\[2mm]

${}^e$ \emph{Grupo de Investigación en Física Teórica} -- GIFT. \\
Universidad Cat\'{o}lica De La Sant\'{i}sima Concepci\'{o}n, Concepción, Chile. \\[2mm]

${}^f$ Instituto de Alta Investigación, Universidad de Tarapacá. \\
Casilla 7D, Arica, Chile. \\[2mm]

\vspace{0.1cm}
${}^*$ {\small{lingua@kth.se}} \qquad \quad 
${}^\dagger$ {\small{dimolina@unap.cl}} \qquad \quad ${}^\star$ {\small{lucrezia.ravera@polito.it}} \qquad \quad ${}^\ddagger$ {\small{ssalgador@gestion.uta.cl}}
\end{center}

\medskip

\begin{abstract}
We present an algebraic method to derive the structure at the basis of the mapping of bosonic algebras of creation and annihilation operators into fermionic algebras, and vice versa, introducing a suitable identification between bosonic and fermionic generators.
The algebraic structure thus obtained corresponds to a deformed Grassmann algebra, involving  anticommuting Grassmann-type variables. The role played by the latter in the implementation of gauge invariance in second quantization within our procedure is then discussed, together with the application of the mapping to the case of the bosonic and fermionic harmonic oscillator Hamiltonians.
\end{abstract}


\vspace{-5mm}

\tableofcontents

\bigskip

\section{Introduction}\label{Introduction}

Bosons and fermions have different characteristics and are distinguished by their intrinsic properties, particularly their spin and statistics. Bosons have integer spin, obey Bose-Einstein statistics, meaning that multiple bosons can occupy the same quantum state simultaneously -- which leads to phenomena such as Bose-Einstein condensation and the behavior of photons in a laser. Fermions obey Fermi-Dirac statistics and are subject to the Pauli exclusion principle, which states that no two fermions can occupy the same quantum state simultaneously.

In quantum field theory (QFT), bosons are typically associated with fields that mediate interactions, fermions are the building blocks of matter. In particle and nuclear physics, composite particles can be bosons if their constituent fermions pair up in such a way that their total spin is an integer (e.g., mesons, made of one quark and one antiquark), while composite particles can be fermions if their constituent fermions combine to give a half-integer spin (e.g., baryons, such as protons and neutrons, which are made of three quarks).
Despite their differences, systems and theories of bosons and fermions can be mapped into each other and one can write, when certain conditions are met and in well-defined regimes, bosons in terms of fermions, and vice versa, allowing a boson-fermion correspondence. For example, the bosonization technique \cite{Coleman:1974bu,emery1979highly,Affleck:1988mua,stone1994bosonization,giamarchi2003quantum,gogolin2004bosonization} is a well-established analytic tool for investigating the low energy regime of one-dimensional interacting fermionic systems, essentially consisting in linearizing the spectrum around the Fermi points, passing to the continuum limit, and finally expressing the fermionic operators in terms of bosonic fields. 
On the other hand, with the so-called fermionization technique, involving Jordan-Wigner transformations \cite{1928ZPhy...47..631J,LIEB1961407}, it is possible to map spin and bosonic systems into fermionic ones -- see \cite{Rodrigues:2016rzw} for a review on bosonization and fermionization.
Furthermore, the Fock space for fermion fields can be identified with the Fock space for boson fields, provided the overall numbers of internal degrees of freedom (d.o.f.) are the same. As a consequence, the respective free field Hamiltonian systems are equivalent, or, as also frequently said, dual. 
The underlying principles of connecting bosonic and fermionic descriptions through dualities are applicable across various domains in QFT, see e.g. the pivotal work \cite{Seiberg:2016gmd}.
Another possible way to relate bosons and fermions may be dictated by supersymmetry. One of the mathematical tools used to formulate and understand supersymmetry is Grassmann-type variables. In particular, supersymmetric theories can be formulated in a way that extends the concept of spacetime into a new structure called superspace, which includes both the usual spacetime coordinates and (anticommuting) Grassmann coordinates.

In this work, we present an algebraic approach to the mapping of the algebra of bosonic creation and annihilation operators (second quantization) into the algebra of fermionic operators, and vice versa, which allows to systematically derive the algebraic structure underlying the mapping. Our procedure is based on the introduction of a proper identification criterion between bosonic and fermionic generators, inherited from a Lie algebra expansion method, called $S\!$-expansion \cite{Izaurieta:2006zz}, and adapted to our purposes.
The basis of the $S\!$-expansion consists in combining the inner multiplication law of a semigroup $S$ with the structure constants of a Lie algebra $\mathcal{G}$. The new Lie algebra thus obtained is called an $S\!$-expanded algebra. From the physical point of view, several theories have been extensively studied using the $S\!$-expansion method, enabling numerous results over recent years, especially in the context of gravitational theories -- see e.g. \cite{Concha:2013uhq,Salgado:2013eut,Concha:2014vka,Concha:2014zsa,Concha:2016hbt,Concha:2016kdz,Gonzalez:2016xwo,Durka:2016eun,Concha:2016tms,Ipinza:2016bfc,Penafiel:2016ufo,Caroca:2017onr,Caroca:2017izc,Concha:2018zeb,Concha:2018jjj,Bergshoeff:2019ctr,Concha:2019lhn,Durka:2019guk,Penafiel:2019czp,Concha:2019dqs,Concha:2020ebl,Concha:2021jnn,Concha:2023bly}.
The identification criterion we adopt here is reminiscent of the one described in \cite{Ipinza:2016bfc}. However, the algebraic structure we obtain is not that of a semigroup. Rather, it corresponds to a graded (deformed) Grassmann algebra, involving anticommuting variables.
It is here derived from the analysis of the (anti)commutation relations of the bosonic and fermionic algebras. 

The reminder of the paper is organised as follows: In Section \ref{Review} we briefly review the identification criterion of the $S\!$-expansion. This criterion is then adapted to write the fermionic generators in terms of bosonic ones -- converting, via the elements of the algebraic structure involved, the bosonic generators into fermionic ones -- and vice versa. In Section \ref{Method} we describe our procedure and derive the algebraic structure underlying the mapping, both in the case in which the identification considered preserve the creation and annihilation operations and in the case in which bosonic/fermionic creation operators are mapped into fermionic/bosonic annihilation operators. 
Moreover, we discuss the role played by the Grassmann-like variables underlying the mapping in realising gauge invariance 
in second quantization, within our procedure. In Section \ref{harmosc} we provide the application to the case of the bosonic and fermionic harmonic oscillator Hamiltonians.
Section \ref{Conclusion} is devoted to the conclusion. 

\section{Review of the identification criterion}\label{Review}

The basis of the $S\!$-expansion consists in combining the multiplication law of a semigroup $S$ with the structure constants of a Lie algebra $\mathcal{G}$
\cite{Izaurieta:2006zz}. The new (larger) Lie algebra obtained through this procedure is named $S\!$-expanded algebra and can be written as $\mathcal{G}_S= S \times \mathcal{G}$. 

In \cite{Ipinza:2016bfc} it was developed an analytic method to find the multiplication table(s) of the set(s) involved in the $S\!$-expansion for reaching a target Lie algebra from a starting one, after having properly chosen the partition over subspaces of the considered algebras (see also \cite{Penafiel:2016ufo}). Let us briefly review the procedure.

In order to derive the multiplication table of the set $S$, the following identification criterion between the $S\!$-expanded generators of the initial Lie algebra and the generators of the target one is introduced: 
\begin{align}\label{identification}
\tilde{T}_A = T_{A,\alpha} \defeq \lambda_\alpha  T_A,
\end{align}
where $T_A$ are the generators in the subspace $V_A$ of the starting algebra, $\tilde{T}_A$ are the generators in the subspace $\tilde{V}_A$ of the target algebra, and where $\lambda_\alpha \in S$ is an element of the set $S$ -- whose subsets are taken, within the $S\!$-expansion method, to be in resonance \cite{Izaurieta:2006zz,Ipinza:2016bfc} with the $\mathcal{G}$-partition in subspaces.
One has to perform the identification \eqref{identification} for each element of the set $S$, associating each element of each subset with the generators in the subspace related to the considered subset.
The whole procedure of association and identification does not affect the internal structure of the generators of the starting algebra.

Under the identification \eqref{identification}, the commutation relations between the generators of the target algebras are linked with the commutation relations of the $S\!$-expanded ones and, factorising the elements of the set $S$, the multiplication relations between these elements are fixed.
For the target algebra one writes 
\begin{align}\label{commtarg}
\left[ \tilde{T}_A, \tilde{T}_B \right] =\tilde{C}_{AB}^{\;\;\;\;C} \tilde{T}_C,
\end{align}
where $\tilde{T}_A$, $\tilde{T}_B$ , and $\tilde{T}_C$ are the generators in the subspaces $\tilde{V}_A$, $\tilde{V}_B$, and $\tilde{V}_C$ of the partition over the target Lie algebra, respectively. With $\tilde{C}_{AB}^{\;\;\;\;C}$ we denote the structure constants of the target Lie algebra.
For the starting algebra we have
\begin{align}\label{commstart}
\left[T_A, T_B \right]= C_{AB}^{\;\;\;\;C}\; T_C ,
\end{align}
with $ C_{AB}^{\;\;\;\;C}$ the corresponding structure constants. 
Then, the commutation relations of the expanded algebra are written in terms of the generators of the starting algebra:
\begin{align}
\left[T_{(A,\alpha)},T_{(B,\beta)}\right]= K_{\alpha \beta}^{\;\;\;\; \gamma} C_{AB}^{\;\;\;\;C}\; T_{(C,\gamma)}, 
\end{align}
or
\begin{align}
    \left[\lambda_\alpha  T_A,\lambda_\beta  T_B \right] = K_{\alpha \beta}^{\;\;\;\; \gamma} C_{AB}^{\;\;\;\;C}\; \lambda_\gamma  T_C,
\end{align}
where the $\lambda$'s are the elements of the set $S$  and where $K_{\alpha \beta}^{\;\;\;\; \gamma}$ is the so-called two-selector, defined as
\begin{equation}\label{kseldef}
K_{\alpha \beta}^{\;\;\;\; \gamma} = \left\{ \begin{aligned} &
\, 1 , \;\;\;\;\; \text{when} \; \lambda_\alpha \cdot \lambda_\beta = \lambda_\gamma,
\\ & \, 0 , \;\;\;\;\; \text{otherwise}. \end{aligned} 
\right.
\end{equation}
One can now write the structure constants of the target algebra in terms of the two-selector and of the structure constants of the starting one:
\begin{align}
\tilde{C}_{AB}^{\;\;\;\;C} := C_{(A,\alpha)(B,\beta)}^{\;\;\;\;\;\;\;\;\;\;\;\;\;\;\;\;(C,\gamma)}= K_{\alpha \beta}^{\;\;\;\gamma}C_{AB}^{\;\;\;\;C}.
\end{align}
By exploiting the identification \eqref{identification}, one can now write the commutation relations of the target algebra \eqref{commtarg} in terms of the commutation relations between the $S\!$-expanded generators of the starting one, factorising the elements of the set $S$ out of the commutators. 
In this way, the following relations are obtained:
\begin{align}\label{commrelafterid}
    (\lambda_\alpha \cdot \lambda_\beta ) \left[T_A, T_B \right] =K_{\alpha \beta}^{\;\;\;\; \gamma} C_{AB}^{\;\;\;\;C} \lambda_\gamma  T_C.
\end{align}
Comparing the commutation relations \eqref{commrelafterid} with the ones of the starting algebra in \eqref{commstart}, one finds
\begin{align}
\lambda_\alpha \cdot \lambda_ \beta = \lambda_\gamma .
\end{align}
Repetition of this procedure for all the commutation rules of the target algebra yields the multiplication rules between the elements of the set $S$, that is its multiplication table. 

In the following, we adopt and adapt the above identification criterion and perform an analogous procedure in order to derive the algebraic structure underlying the mapping between bosonic and fermionic algebras.

\section{Algebraic approach to the boson-fermion mapping}\label{Method}

We consider bosonic and fermionic creation and annihilation operators acting on modes, e.g., sites of a lattice. For the sake of clarity, we label bosonic modes with $I,J,\ldots=1,\ldots,N_B$ and fermionic modes with $i,j,\ldots = 1, \ldots, N_F$.

We consider the bosonic algebra $\mathcal{G}_B$ generated by $\lbrace{a_I,a^\dagger_I\rbrace}$, being $a_I$ and $a^\dagger_I$ the bosonic annihilation and creation operators, respectively, acting on the mode $I$. They satisfy the following commutation relations:
\begin{equation}
\begin{aligned}
&[a_I,a_J ^\dagger] = a_I a_J ^\dagger - a_J ^\dagger a_I = \delta_{IJ} ,  \\
&[a_I,a_J]=[a_I ^\dagger ,a_J ^\dagger]=0.\label{bos_alg}
\end{aligned}
\end{equation}
As target algebra, we consider the fermionic algebra $\mathcal{G}_F$ generated by $\lbrace{ c_i,c^\dagger_i\rbrace}$, where $c_i$ and $c^\dagger_i$ are fermionic annihilation and creation operators, respectively, acting on the mode $i$. These generators satisfy the following anticommutation relations:
\begin{equation}
\begin{aligned}
&\{c_i,c_j^\dagger\}= \delta_{ij} , \\
&\{c_i,c_j\}=\{c_i^\dagger, c_j ^\dagger\}=0. \label{fer_alg}
\end{aligned}
\end{equation}

Now, our goal is to find the algebraic structure linking the bosonic algebra $\mathcal{G}_B$ and the fermionic algebra $\mathcal{G}_F$. To this aim, we apply the identification criterion developed in the $S\!$-expansion context, adapted to our framework.
As we shall see, two main different identifications are possible.

\subsection{Mapping that preserves the creation/annihilation operation}\label{presmap}

We write the following identification relations between the fermionic and bosonic generators:
\begin{equation}\label{part}
\left\{
\begin{aligned}
&\,c_i={\lambda_{i}}^I  a_I, \\
&\,c_i^\dagger = {\lambda_i}^{I\dagger}  a_I^\dagger , 
\end{aligned} \right.
\end{equation}
where ${\lambda_{i}}^I$ and ${\lambda_i}^{I\dagger}$ are the elements of a set $\Lambda$. Notice that the $\lambda$'s carry mode labels, and the contracted indices in \eqref{part} are summed over (summation is implied).

We extract the algebraic properties of the set $\Lambda$ involved in the mapping -- which, at this stage, corresponds to consistently writing, using the identification \eqref{part}, fermions in terms of bosons -- by analysing the (anti)commutation relations of the fermionic and of the bosonic algebras. 
We start by considering the first anticommutation relation in \eqref{fer_alg}.
Substituting the expressions in \eqref{part} for the fermionic generators $c_i$ and $c_j^\dagger$, we obtain
\begin{equation}
\begin{aligned}
\{c_i,c_j^\dagger\} & = \delta_{ij} \\
& = \{{\lambda_i}^I  a_I , {\lambda_j}^{J\dagger}  a_J^\dagger \}= ({\lambda_i}^I \cdot {\lambda_j}^{J \dagger}) a_I a_J^{\dagger} + ({\lambda_j}^{J\dagger} \cdot {\lambda_i}^I ) a_J^{\dagger}a_I .
\end{aligned}
\end{equation}
Now we see that, requiring, as a consistency relation to be fulfilled by the $\lambda$'s,
\begin{align}\label{rel1}
{\lambda_i}^I \cdot {\lambda_j}^{J\dagger} = -{\lambda_j}^{J\dagger} \cdot {\lambda_i}^I ,
\end{align}
we get
\begin{equation}
\begin{aligned}
\{c_i,c_j^\dagger\}&=\delta_{ij}\\
&= {\lambda_i}^I \cdot {\lambda_j}^{J\dagger} (a_I a_J^{\dagger} - a_J^{\dagger}a_I)= {\lambda_i}^I \cdot {\lambda_j}^{J\dagger}  [a_I , a_J^{\dagger}] = {\lambda_i}^I \cdot {\lambda_j}^{J\dagger}  \delta_{IJ} ,
\end{aligned}
\end{equation}
where in the last step we have used the first commutation relation in \eqref{bos_alg}. Hence, we are left with the following relation:
\begin{align}
\label{deformGr1}
{\lambda_i}^I \cdot \lambda_{jI}^{\dagger} = \delta_{ij}.
\end{align}
On the other hand, taking the anticommutation relation $\{c_i,c_j\}=0$
and using the identification \eqref{part}, we can write
\begin{equation}
\begin{aligned}
\{c_i,c_j\} &=0\\
&= \{{\lambda_i}^I  a_I, {\lambda_j}^J a_J\} ={\lambda_i}^I \cdot{\lambda_j}^J  [a_I, a_J] = 0,
\end{aligned}
\end{equation}
where we have also introduced and exploited the consistency requirement
\begin{align}\label{rel2}
{\lambda_i}^I \cdot {\lambda_j}^J = -{\lambda_j}^J \cdot {\lambda_i}^I 
\end{align}
and used the commutation relation $[a_I,a_J]=0$.
Analogously, considering the anticommutation relation $\{c_i^\dagger , c_j^\dagger \}=0$, using the bosonic commutation relation $[a_I^\dagger,a_J^\dagger]=0$, and implementing the identification \eqref{part}
we get 
\begin{align}\label{rel3}
{\lambda_i}^{I\dagger} \cdot {\lambda_j}^{J\dagger} = -{\lambda_j}^{J\dagger} \cdot {\lambda_i}^{I\dagger} .
\end{align}
Hence, we end up with the multiplication rules \eqref{rel1}, \eqref{deformGr1}, \eqref{rel2}, and \eqref{rel3} between the elements of $\Lambda$.
The algebraic structure obtained is thus
\begin{equation}
\begin{aligned}
& \{ {\lambda_i}^I , {\lambda_j}^{J\dagger} \} = {\lambda_i}^I \cdot {\lambda_j}^{J\dagger} + {\lambda_j}^{J\dagger} \cdot {\lambda_i}^I =  0 , \\
& \{ {\lambda_i}^I , {\lambda_j}^J \} =  {\lambda_i}^I \cdot {\lambda_j}^J + {\lambda_j}^J  \cdot  {\lambda_i}^I  = 0 , \\
& \{ {\lambda_i}^{I\dagger}, {\lambda_j}^{J\dagger} \} = {\lambda_i}^{I\dagger} \cdot {\lambda_j}^{J\dagger} + {\lambda_j}^{J\dagger} \cdot {\lambda_i}^{I\dagger} = 0 ,
\end{aligned}
\end{equation}
which corresponds to a graded Grassmann algebra, involving anticommuting Grassmann-type variables.\footnote{The anticommuting elements ${\lambda_i}^I$ and ${\lambda_i}^{I\dagger}$ are $\mathbb{Z}_2$ odd-graded (fermionic) elements.
The algebra they generate is $\mathbb{Z}_2$-graded.} Furthermore, one may call it a ``deformed" Grassmann algebra, given that the $\lambda,\lambda^\dagger$'s obey the multiplication rule \eqref{deformGr1} -- which can be seen as an extra condition with respect to those appearing in a standard Grassmann algebra.
Note that Pauli exclusion principle on fermions is still naturally satisfied -- and it will actually be so also in the other mappings we will present, due to the fact that the original (anti)commutation relations are respected.

\paragraph{Inverse mapping of fermionic into bosonic operators}

One can also consider the inverse mapping of fermions into bosons, by applying an analogous procedure, i.e. by writing an identification
\begin{equation}\label{parta}
\left\{
\begin{aligned}
&\,a_I={\lambda_I}^i  c_i, \\
&\,a_I^\dagger ={\lambda_I}^{i \dagger}  c_i^\dagger , 
\end{aligned} \right.
\end{equation}
with ${\lambda_I}^i$ and ${\lambda_I}^{i \dagger}$ the elements of the set involved in the mapping. In this case, from the analysis of the first commutation relation in \eqref{bos_alg} we get
\begin{equation}\label{invmap1}
\begin{aligned}
[a_I,a_J ^\dagger] & = \delta_{IJ} \\
& = [{\lambda_I}^i  c_i , {\lambda_J}^{j\dagger}  c_j^\dagger ] = ({\lambda_I}^i \cdot {\lambda_J}^{j \dagger}) c_i c_j^{\dagger} - ({\lambda_J}^{j\dagger} \cdot {\lambda_I}^i )  c_j^{\dagger}c_i \\ 
& = {\lambda_I}^i \cdot {\lambda_J}^{j\dagger}  (c_i c_j^{\dagger} + c_j^{\dagger}c_i)= {\lambda_I}^i \cdot {\lambda_J}^{j\dagger}  \{c_i , c_j^{\dagger}\} =  {\lambda_I}^i \cdot {\lambda_J}^{j\dagger} \delta_{ij}, 
\end{aligned}
\end{equation}
where in the second line we have implemented the consistency requirement
\begin{align}\label{rel1a}
{\lambda_I}^i \cdot {\lambda_J}^{j\dagger} = -{\lambda_J}^{j\dagger} \cdot {\lambda_I}^i ,
\end{align}
and made use of the anticommutation relation $\{c_i,c_j\}=\delta_{ij}$. Hence, from \eqref{invmap1} we also get
\begin{align}\label{deformGr1a}
    {\lambda_I}^i \cdot \lambda_{Ji}^{\dagger} = \delta_{IJ}.
\end{align}
On the other hand, we have
\begin{equation}
\begin{aligned}
[a_I,a_J] & = 0 \\
& = [{\lambda_I}^i  c_i , {\lambda_J}^{j}  c_j ] = ({\lambda_I}^i \cdot {\lambda_J}^{j}) c_i c_j - ({\lambda_J}^{j} \cdot {\lambda_I}^i )  c_j c_i  \\ 
& = {\lambda_I}^i \cdot {\lambda_J}^{j} (c_i c_j + c_j c_i)= {\lambda_I}^i \cdot {\lambda_J}^{j}  \{c_i , c_j \} = 0 ,
\end{aligned}
\end{equation}
where we have used, for consistency,
\begin{align}\label{rel2a}
{\lambda_I}^i \cdot {\lambda_J}^{j} = -{\lambda_J}^{j} \cdot {\lambda_I}^i .
\end{align}
Analogously, from the analysis of  $[a^\dagger_I,a^\dagger_J] = 0$, via $ \{c^\dagger_i , c^\dagger_j \} = 0$, we can derive, for consistency, the multiplication rule
\begin{align}\label{rel3a}
{\lambda_I}^{i\dagger} \cdot {\lambda_J}^{j\dagger} = -{\lambda_J}^{j\dagger} \cdot {\lambda_I}^{i\dagger} .
\end{align}
Therefore, in the case of the inverse mapping we end up with the multiplication rules \eqref{rel1a}, \eqref{deformGr1a}, \eqref{rel2a}, and \eqref{rel3a} between the elements of the set underlying the mapping.
Also in this case, the algebraic structure obtained corresponds to a graded, deformed Grassmann algebra, involving anticommuting Grassmann-type variables, and reads
\begin{equation}
\begin{aligned}
& \{ {\lambda_I}^i , {\lambda_J}^{j\dagger} \} = 0 , \\
& \{ {\lambda_I}^i , {\lambda_J}^j \} = 0 , \\
& \{ {\lambda_I}^{i\dagger}, {\lambda_J}^{j\dagger} \} = 0 ,
\end{aligned}
\end{equation}
together with the extra relation \eqref{deformGr1a}.

\medskip
Let us observe that, considering the identifications \eqref{part} and \eqref{parta} together, 
one may then also write 
\begin{align}
    a_J = {\lambda_J}^i  c_i = {\lambda_J}^i   {\lambda_i}^I  a_I  = \left({\lambda_J}^i \cdot {\lambda_i}^I \right)  a_I \quad \Rightarrow \quad & {\lambda_J}^i \cdot {\lambda_i}^I = {\delta_J}^I , \label{ll1a} \\
    a^\dagger_J = {\lambda_J}^{i\dagger}  c^\dagger_i = {\lambda_J}^{i\dagger}  {\lambda_i}^{I\dagger} a^\dagger_I  = \left({\lambda_J}^{i\dagger}  \cdot {\lambda_i}^{I\dagger}  \right)  a^\dagger_I \quad \Rightarrow \quad & {\lambda_J}^{i\dagger}  \cdot {\lambda_i}^{I\dagger}  = {\delta_J}^I , \label{ll2a} \\
    c_j = {\lambda_j}^I  a_I =  {\lambda_j}^I  {\lambda_I}^i  c_i = \left(  {\lambda_j}^I  \cdot {\lambda_I}^i \right)  c_i \quad \Rightarrow \quad &  {\lambda_j}^I \cdot {\lambda_I}^i = {\delta_j}^i , \label{ll1c} \\
    c^\dagger_j = {\lambda_j}^{I\dagger}  a^\dagger_I =  {\lambda_j}^{I\dagger} {\lambda_I}^{i\dagger} c^\dagger_i  = \left( {\lambda_j}^{I\dagger} \cdot {\lambda_I}^{i\dagger} \right)  c^\dagger_i \quad \Rightarrow \quad & {\lambda_j}^{I\dagger} \cdot {\lambda_I}^{i\dagger}= {\delta_j}^i , \label{ll2c} 
\end{align}
ending up with (multiplicative inverse) relations between the ${\lambda_I}^i,{\lambda_I}^{i\dagger}$ and the ${\lambda_i}^I,{\lambda_i}^{I\dagger}$ elements.

\subsection{Mapping that exchanges the creation and annihilation operations}\label{nonpresmap}

Another identification that one may consider, in contrast to \eqref{part}, is the following:
\begin{equation}\label{part1}
\left\{
\begin{aligned}
&\,c_i={\uplambda_{i}}^{I\dagger}  a^\dagger_I, \\
&\,c_i^\dagger ={\uplambda_i}^{I}  a_I , 
\end{aligned} \right.
\end{equation}
where the creation/annihilation of bosons is translated into the corresponding annihilation/creation of fermions. 

From the study of the first anticommutation relation in \eqref{fer_alg} we obtain
\begin{equation}
\begin{aligned}
\{c_i,c_j^\dagger\} & = \delta_{ij} \\
& = \{{\uplambda_i}^{I\dagger}  a_I^\dagger , {\uplambda_j}^{J}  a_J \}= ({\uplambda_i}^{I\dagger} \cdot {\uplambda_j}^{J}) a_I^\dagger a_J  + ({\uplambda_j}^{J} \cdot {\uplambda_i}^{I\dagger} )  a_J a_I^{\dagger}  \\
& = {\uplambda_i}^{I\dagger} \cdot {\uplambda_j}^{J} (a_I^{\dagger} a_J - a_J a_I^{\dagger})=-{\uplambda_i}^{I\dagger} \cdot {\uplambda_j}^{J} [a_J , a_I^{\dagger}] = {\uplambda_j}^J \cdot {\uplambda_i}^{I\dagger}  \delta_{JI} ,
\end{aligned}
\end{equation}
where we have required, for consistency, the relation 
\begin{align}\label{relmnp}
    {\uplambda_i}^{I\dagger} \cdot {\uplambda_j}^{J} = - {\uplambda_j}^{J} \cdot {\uplambda_i}^{I\dagger}
\end{align}
to hold (which is reminiscent of the relation \eqref{rel1} previously obtained for the $\lambda$'s and $\lambda^\dagger$'s elements),
and where we have also used the first commutation relation in \eqref{bos_alg}. Therefore, we end up with the relation 
\begin{align}\label{deformGr1mnp}
    {\uplambda_j}^I \cdot {\uplambda_{iI}}^{\dagger} = \delta_{ij}
\end{align}
between the $\uplambda$'s and $\uplambda^\dagger$'s (analogous to the relation \eqref{deformGr1} previously obtained for the $\lambda$'s and $\lambda^\dagger$'s). 
Then, considering the other anticommutation relations in the fermionic algebra we get
\begin{equation}
\begin{aligned}
\{c_i,c_j\} &=0\\
&= \{{\uplambda_i}^{I\dagger}  a^\dagger_I, {\uplambda_j}^{J\dagger}  a^\dagger_J\} ={\uplambda_i}^{I\dagger} \cdot {\uplambda_j}^{J\dagger} [a^\dagger_I, a^\dagger_J] = 0
\end{aligned}
\end{equation}
and
\begin{equation}
\begin{aligned}
\{c^\dagger_i,c^\dagger_j\} &=0\\
&= \{{\uplambda_i}^{I}  a_I, {\uplambda_j}^{J}  a_J\} ={\uplambda_i}^{I} \cdot {\uplambda_j}^{J} [a_I, a_J] = 0,
\end{aligned}
\end{equation}
where we have also introduced and exploited the consistency requirements
\begin{align}\label{rel3mnp}
    {\uplambda_i}^{I\dagger} \cdot {\uplambda_j}^{J\dagger} = -{\uplambda_j}^{J\dagger} \cdot {\uplambda_i}^{I\dagger} 
\end{align}
and
\begin{align}\label{rel2mnp}
    {\uplambda_i}^I \cdot {\uplambda_j}^J = -{\uplambda_j}^J \cdot {\uplambda_i}^I ,
\end{align}
respectively analogous to \eqref{rel3} and \eqref{rel2} previously obtained for the $\lambda^\dagger$'s and $\lambda$'s.
Hence, the algebraic structure underlying the mapping that converts the creation operation into the annihilation one and vice versa is the same we have obtained in the case of the mapping preserving the creation/annihilation operation, namely a (deformed, due to \eqref{deformGr1mnp}) Grassmann algebra involving anticommuting Grassmann-type variables.

\paragraph{Inverse mapping of fermionic into bosonic operators}

In a similar way, we may now introduce the following identification:
\begin{equation}\label{part1a}
\left\{
\begin{aligned}
&\,a_I={\uplambda_{I}}^{i\dagger}  c^\dagger_i, \\
&\,a_I^\dagger ={\uplambda_I}^{i}  c_i , 
\end{aligned} \right.
\end{equation}
translating the creation/annihilation of fermions into the annihilation/creation of bosons -- which is the inverse mapping of \eqref{part1}. Thus, from the analysis of the bosonic algebra we get
\begin{equation}\label{invmap1a}
\begin{aligned}
[a_I,a_J ^\dagger] & = \delta_{IJ} \\
& = [{\uplambda_I}^{i\dagger}  c^\dagger_i , {\uplambda_J}^{j}  c_j ] = ({\uplambda_I}^{i\dagger} \cdot {\uplambda_J}^{j}) c^\dagger_i c_j - ({\uplambda_J}^{j} \cdot {\uplambda_I}^{i\dagger} )  c_j c^\dagger_i \\ 
& = {\uplambda_I}^{i \dagger} \cdot {\uplambda_J}^{j} (c_j c_i^{\dagger} + c_i^{\dagger}c_j)= {\uplambda_I}^{i\dagger} \cdot {\uplambda_J}^{j}  \{c_j , c_i^{\dagger}\} = {\uplambda_I}^{i\dagger} \cdot {\uplambda_J}^{j}  \delta_{ji}, 
\end{aligned}
\end{equation}
where we have also implemented the consistency requirement 
\begin{align}\label{rel1amnp}
{\uplambda_I}^i \cdot {\uplambda_J}^{j\dagger} = -{\uplambda_J}^{j\dagger} \cdot {\uplambda_I}^i ,
\end{align}
analogous to the relation \eqref{rel1a} previously written for the $\lambda$'s and $\lambda^\dagger$'s. Therefore, from \eqref{invmap1a} we are left with the relation
\begin{align}\label{deformGr2a}
    {\uplambda_I}^{i\dagger} \cdot {\uplambda}_{Ji} = \delta_{IJ} \quad \Rightarrow \quad \uplambda_{Ji} \cdot {\uplambda_I}^{i\dagger}   = - \delta_{IJ}.
\end{align}
Moreover, we have
\begin{equation}
\begin{aligned}
[a_I,a_J] & = 0 \\
& = [{\uplambda_I}^{i\dagger}  c^\dagger_i , {\uplambda_J}^{j\dagger}  c^\dagger_j ] = ({\uplambda_I}^{i\dagger} \cdot {\uplambda_J}^{j\dagger}) c^\dagger_i c^\dagger_j - ({\uplambda_J}^{j\dagger} \cdot {\uplambda_I}^{i\dagger} )  c^\dagger_j c^\dagger_i  \\ 
& = {\uplambda_I}^{i\dagger} \cdot {\uplambda_J}^{j\dagger} (c^\dagger_i c^\dagger_j + c^\dagger_j c^\dagger_i)={\uplambda_I}^{i\dagger} \cdot {\uplambda_J}^{j\dagger}  \{c^\dagger_i , c^\dagger_j \} = 0 ,
\end{aligned}
\end{equation}
where we have used 
\begin{align}\label{rel3amnp}
{\uplambda_I}^{i\dagger} \cdot {\uplambda_J}^{j\dagger} = -{\uplambda_J}^{j\dagger} \cdot {\uplambda_I}^{i\dagger} ,
\end{align}
reminiscent of \eqref{rel3a} among the $\lambda^\dagger$'s.
Analogously, from the analysis of  $[a^\dagger_I,a^\dagger_J] = 0$, via $ \{c_i , c_j \} = 0$, we derive the multiplication rule
\begin{align}\label{rel2amnp}
    {\uplambda_I}^i \cdot {\uplambda_J}^{j} = -{\uplambda_J}^{j} \cdot {\uplambda_I}^i ,
\end{align}
reminiscent of \eqref{rel2a} for the $\lambda$'s.
Hence, the algebraic structure obtained for the $\uplambda$'s and $\uplambda^\dagger$'s is, again, a graded (deformed) Grassmann algebra, with the deformation due to the extra relation \eqref{deformGr2a}.

\medskip
Notice that, as previously observed in the case of the mapping that preserves the creation/annihilation operation, one may now consider \eqref{part1} and \eqref{part1a} together and write
\begin{align}
    a_J = {\uplambda_J}^{i\dagger}  c^\dagger_i =  {\uplambda_J}^{i\dagger}  {\uplambda_i}^I  a_I  = \left({\uplambda_J}^{i\dagger}  \cdot {\uplambda_i}^I\right)  a_I \quad \Rightarrow \quad & {\uplambda_J}^{i\dagger}  \cdot {\uplambda_i}^I= {\delta_J}^I , \label{ull1a} \\
    a^\dagger_J = {\uplambda_J}^{i}  c_i = {\uplambda_J}^{i} {\uplambda_i}^{I\dagger} a^\dagger_I  = \left({\uplambda_J}^{i} \cdot {\uplambda_i}^{I\dagger} \right)  a^\dagger_I \quad \Rightarrow \quad & {\uplambda_J}^{i} \cdot {\uplambda_i}^{I\dagger} = {\delta_J}^I , \label{ull2a} \\ 
    c_j = {\uplambda_j}^{I\dagger}  a^\dagger_I =   {\uplambda_j}^{I\dagger}  {\uplambda_I}^i  c_i = \left( {\uplambda_j}^{I\dagger} \cdot {\uplambda_I}^i \right)  c_i \quad \Rightarrow \quad & {\uplambda_j}^{I\dagger} \cdot {\uplambda_I}^i = {\delta_j}^i , \label{ull1c} \\
    c^\dagger_j = {\uplambda_j}^{I}  a_I =  {\uplambda_j}^{I}  {\uplambda_I}^{i\dagger}  c^\dagger_i  = \left({\uplambda_j}^{I}  \cdot {\uplambda_I}^{i\dagger}  \right)  c^\dagger_i \quad \Rightarrow \quad & {\uplambda_j}^{I}  \cdot {\uplambda_I}^{i\dagger} = {\delta_j}^i , \label{ull2c} 
\end{align}
obtaining (multiplicative inverse) relations between the ${\uplambda_I}^i,{\uplambda_I}^{i\dagger}$ and the ${\uplambda_i}^I,{\uplambda_i}^{I\dagger}$ elements.

\subsection{Gauge invariance in second quantization and role of the Grassmann-type variables}

Let us conclude this section with an observation on gauge invariance and symmetry reduction in our context. In second quantization, gauge transformations correspond to phase transformations of the creation and annihilation operators for fermions/bosons. These transformations reflect the invariance of the system under local or global phase changes -- in particular, the latter leave physical observables, such as occupation numbers, unchanged.
For fermionic creation and annihilation operators, for a global $\mathrm{U}(1)$ gauge transformation the operators transform as 
\begin{align}
    c_i \rightarrow c'_i = e^{i\theta} c_i , \quad
    c_i^\dagger \rightarrow {c'}_i^\dagger = e^{-i\theta} c_i^\dagger ,
\end{align}
where $\theta$ is the phase angle of the transformation. This kind of transformation corresponds to a global symmetry where the phase factor is the same for all operators.
Analogously, for bosonic creation and annihilation operators one may write
\begin{align}
    a_I \rightarrow a'_I = e^{i\theta} a_I , \quad
    a_I^\dagger \rightarrow {a'}_I^\dagger = e^{-i\theta} a_I^\dagger .
\end{align}
On the other hand, for a \emph{local} $\mathrm{U}(1)$ gauge transformation the phase factor can depend on ``position" (on the mode index, e.g. site of a lattice model). While global gauge transformations leave physical observables unchanged because they apply a constant phase to all states in the system, local gauge transformations can affect observables unless they are compensated by other modifications in the system (such as the introduction of gauge fields). Let us denote the aforementioned position or mode dependence by $\theta_i$ (or $\theta_I$), reflecting a local symmetry.
The fermionic creation and annihilation operators then transform as
\begin{align}\label{gaugetrf}
c_i \rightarrow c'_i = e^{i\theta_i} c_i , \quad
c_i^\dagger \rightarrow {c'}_i^\dagger = e^{-i\theta_i} c_i^\dagger ,
\end{align}
while for bosons we have
\begin{align}\label{gaugetrb}
a_I \rightarrow a'_I = e^{i\theta_I} a_I , \quad
a_I^\dagger \rightarrow {a'}_I^\dagger = e^{-i\theta_I} a_I^\dagger .
\end{align}
Taking this into account and exploiting the identifications \eqref{part} and \eqref{parta} inducing the mappings previously discussed, we find the $\theta_I$ gauge transformations of the different $\lambda$'s and $\lambda^\dagger$'s elements to be
\begin{align}\label{gaugetrlthetaI}
    {\lambda_i}^I \rightarrow  {\lambda'_i}^I = e^{-i\theta_I} {\lambda_i}^I , \quad
    {\lambda_i}^{I\dagger} \rightarrow {\lambda'_i}^{I\dagger} = e^{i\theta_I} {\lambda_i}^{I\dagger} , \quad  {\lambda_I}^i \rightarrow {\lambda'_I}^i = e^{i\theta_I} {\lambda_I}^i , \quad
    {\lambda_I}^{i\dagger} \rightarrow {\lambda'_I}^{i\dagger} = e^{-i\theta_I} {\lambda_I}^{i\dagger} ,
\end{align}
while the $\theta_i$ gauge transformations are
\begin{align}\label{gaugetrlthetai}
    {\lambda_i}^I \rightarrow {\lambda'_i}^I = e^{i\theta_i} {\lambda_i}^I , \quad {\lambda_i}^{I\dagger} \rightarrow {\lambda'_i}^{I\dagger} = e^{-i\theta_i} {\lambda_i}^{I\dagger}, \quad {\lambda_I}^i \rightarrow  {\lambda'_I}^i = e^{-i\theta_i} {\lambda_I}^i , \quad
    {\lambda_I}^{i\dagger} \rightarrow {\lambda'_I}^{i\dagger} = e^{i\theta_i} {\lambda_I}^{i\dagger} .
\end{align}
Correspondingly, the creation and annihilation operators $c^\dagger_i,c_i$ and $a^\dagger_I,a_I$ can be seen as composite objects, invariant under (part of) the gauge symmetry. In other words, one may formally write, omitting mode labels just to lighten the notation, 
\begin{align}
    a^\lambda:=\lambda^{-1}a=\lambda^{-1}(\lambda c)=c , \quad a^{\dagger \lambda^\dagger}:=\lambda^{\dagger -1}a^\dagger=\lambda^{\dagger -1}(\lambda^\dagger c^\dagger)=c^\dagger ,
\end{align}
and, considering the $\lambda$'s and $\lambda^\dagger$'s involved in the inverse mapping, 
\begin{align}
    c^\lambda:=\lambda^{-1}a=\lambda^{-1}(\lambda a)=a , \quad c^{\dagger \lambda^\dagger}:=\lambda^{\dagger -1}a^\dagger=\lambda^{\dagger -1}(\lambda^\dagger a^\dagger)=a^\dagger .
\end{align}
This allows to interpret the creation and annihilation bosonic (fermionic) operators as dressed fermionic (bosonic) ones,\footnote{This can be seen as a \emph{self-dressing}, as it is extracted from the operators themselves.} where the dressing is implemented by the Grassmann-type variables $\lambda,\lambda^\dagger$ -- which one may therefore name \emph{dressing Grassmannian variables}. As shown in \eqref{gaugetrlthetaI}-\eqref{gaugetrlthetai}, under local gauge transformations the latter have their own phase adjusted to cancel the phase change in the bosonic/fermionic operators, thus creating operators that are gauge-invariant under the symmetry that has been reduced.
Note that, depending on the type of system being considered -- i.e. whether it is fermionic or bosonic, part of the above mentioned gauge symmetry transformations of the $\lambda,\lambda^\dagger$'s can be interpreted as residual gauge symmetries of the (partially) dressed operators. 

The same arguments above can be applied, in an analogous way, to the mappings that exchanges the creation and annihilation operations. In this case we get, taking into account \eqref{gaugetrf}-\eqref{gaugetrb} and considering the mappings \eqref{part1}/\eqref{part1a},
\begin{align}\label{gaugetruplI}
    {\uplambda_i}^I \rightarrow  {\uplambda'_i}^I = e^{-i\theta_I} {\uplambda_i}^I , \quad
    {\uplambda_i}^{I\dagger} \rightarrow {\uplambda'_i}^{I\dagger} = e^{i\theta_I} {\uplambda_i}^{I\dagger} , \quad  {\uplambda_I}^i \rightarrow {\uplambda'_I}^i = e^{-i\theta_I} {\uplambda_I}^i , \quad
    {\uplambda_I}^{i\dagger} \rightarrow {\uplambda'_I}^{i\dagger} = e^{i\theta_I} {\uplambda_I}^{i\dagger} ,
\end{align}
and
\begin{align}\label{gaugetrupli}
    {\uplambda_i}^I \rightarrow {\uplambda'_i}^I = e^{-i\theta_i} {\uplambda_i}^I , \quad {\uplambda_i}^{I\dagger} \rightarrow {\uplambda'_i}^{I\dagger} = e^{i\theta_i} {\uplambda_i}^{I\dagger}, \quad {\uplambda_I}^i \rightarrow  {\uplambda'_I}^i = e^{-i\theta_i} {\uplambda_I}^i , \quad
    {\uplambda_I}^{i\dagger} \rightarrow {\uplambda'_I}^{i\dagger} = e^{i\theta_i} {\uplambda_I}^{i\dagger} ,
\end{align}
under $\theta_I$ and $\theta_i$ gauge transformations, respectively. Therefore, one may formally write, on the one hand, 
\begin{align}
    a^{\uplambda^\dagger}:=\uplambda^{\dagger -1}a=\uplambda^{\dagger -1}(\uplambda^\dagger c^\dagger)=c^\dagger , \quad a^{\dagger \uplambda}:=\uplambda^{-1}a^\dagger=\uplambda^{-1}(\uplambda c)=c,
\end{align}
and, on the other, 
\begin{align}
    c^{\uplambda^\dagger}:=\uplambda^{\dagger -1}a^\dagger=\lambda^{\dagger -1}(\uplambda^\dagger a^\dagger)=a^\dagger , \quad c^{\dagger \uplambda}:=\uplambda^{-1}a=\uplambda^{-1}(\uplambda a)=a ,
\end{align}
considering the $\uplambda$'s and $\uplambda^\dagger$'s of the inverse mapping.

\section{Application to the Hamiltonians of the bosonic/fermionic harmonic oscillators}\label{harmosc}

We will now apply the algebraic mappings and reasoning presented above to the case of the (free) Hamiltonian of the bosonic and fermionic harmonic oscillators. 
Here we explicitly apply the mapping that preserves the creation/annihilation operation; the same reasoning can be performed for the mapping that exchanges the creation/annihilation operation, leading, in a straightforward way, to algebraically analogous results.
The Hamiltonian of the bosonic harmonic oscillator is
\begin{align}
    H_B = \frac{\hbar\omega}{2}\{a^\dagger_I,a_I\}=\frac{\hbar\omega}{2}( a^\dagger_I a_I + a_I a^\dagger_I  )=\hbar\omega\left(a^\dagger_I a_I + \frac{1}{2} \right),\label{HB}
\end{align}
where summation over $I$ is implied.
Applying the identification \eqref{parta} to $H_B$ and working the way down with the induced algebraic mapping, exploiting also the previously derived relations between the associated variables $\lambda,\lambda^\dagger$'s, we get
\begin{align}
    H_B = \frac{\hbar\omega}{2}  {\lambda_I}^{i\dagger} \cdot {\lambda_I}^j (c^\dagger_i c_j - c_j c^\dagger_i ) = \frac{\hbar\omega}{2}   {\lambda_I}^{i\dagger} \cdot {\lambda_I}^j   [c^\dagger_i,c_j]  . \label{HB_3}
\end{align}
In other words, with our procedure one can consistently derive the Hamiltonian $H_B$ starting from the commutator $ [c^\dagger_i,c_j]$ between fermionic operators, properly building, by means of $\lambda,\lambda^\dagger$, an object that is invariant under $\theta_i$ gauge transformations -- therefore this symmetry is reduced -- and also invariant under $\theta_I$ gauge transformations by construction (as expected for $H_B$).

In a similar way, one can consider the Hamiltonian of the so-called fermionic harmonic oscillator
\begin{align}
    H_F = \frac{\hbar\omega}{2} [c^\dagger_i ,c_i] ,
\end{align}
and write, by applying the identification \eqref{part} and using the relations between the $\lambda,\lambda^\dagger$'s underlying the associated mapping,
\begin{align}
    H_F = \frac{\hbar\omega}{2}{\lambda_i}^{I\dagger} \cdot {\lambda_i}^J ( a^\dagger_I a_J + a_J a^\dagger_I ) = \frac{\hbar\omega}{2} {\lambda_i}^{I\dagger} \cdot {\lambda_i}^J \{a^\dagger_I,a_J\} .
\end{align}
We can therefore see that the Hamiltonian $H_F$ can be derived from the anticommutator $\{a^\dagger_I,a_J\}$ between bosonic operators. This is done by building, via the Grassmann-type variables $\lambda,\lambda^\dagger$'s appearing in \eqref{part}, an object that is invariant under $\theta_I$ gauge transformations. By construction, it is also invariant under $\theta_i$ transformations, as indeed expected for $H_F$.

Under the above perspective, the bosonic (fermionic) Hamiltonian $H_B$ ($H_F$) can be therefore seen as dressed objects, derived from bare commutation (anticommutation) relations between creation and annihilation operators.

\section{Conclusion}\label{Conclusion}

In this work, we have presented an algebraic approach to the mapping of Lie algebras $\mathcal{G}_B$ of bosonic creation and annihilation operators into algebras $\mathcal{G}_F$ of fermionic creation and annihilation operators, and vice versa.

We have introduced a specific identification criterion, inherited from a Lie algebra expansion method -- known as $S\!$-expansion -- and adapted to our purposes, between the bosonic and the fermionic generators. We have then used the (anti)commutation relations of the bosonic and fermionic algebras in order to determine the algebraic structure underlying the mapping.
The latter corresponds to a graded, deformed Grassmann algebra, involving  anticommuting Grassmann-type variables. 
The deformation is due to an extra relation among the elements of the algebra necessary for the consistency of the construction based on the identification criterion adopted case by case. 
We have presented different mappings, all based on a Grassmann-type algebra, that either preserve the creation/annihilation operations or exchange them. In both cases, we have analysed the mapping of bosonic to fermionic operators and its inverse.

We have then discussed the role played by the Grassmann-type variables concerning gauge invariance in second quantization.
We have found that, within our procedure for the various mappings, the bosonic/fermionic creation and annihilation operators can be seen as dressed operators, where the dressing is provided by the Grassmann-type variables underlying the mapping, hence interpretable as dressing Grassmannian variables. 
This kind of objects might also be extracted from the physical content of the model by considering, e.g., the polar decomposition of (complex) eigenvalues, which separates magnitude and phase. 
We leave such analysis to future investigations.

We have also provided an example of application to the case of the Hamiltonians of the bosonic and fermionic harmonic oscillators, showing that they can be seen as dressed objects, derived from the -- bare -- commutation (anticommutation) relations between the fermionic (bosonic) creation and annihilation operators.

Future perspectives include the application/extension of the algebraic method here developed within a field-theoretic context and the analysis of various Hamiltonian systems (e.g., Hubbard models) by means of the here proposed mappings, considering also on-site and in-site interaction terms. 

Finally, it would be interesting to see if and how our approach actually makes contact with supergeometry and/or supersymmetry, in particular concerning the way in which we recover gauge invariance, especially in the context of the so-called \emph{unconventional supersymmetry} (Ususy) \cite{Alvarez:2011gd,Alvarez:2013tga,Guevara:2016rbl,Alvarez:2021zhh}, which has been shown to play a relevant role in
the construction of analogue (supergravity) models, providing a macroscopic description
of the electronic properties of graphene-like materials. There, supersymmetry is not manifest, but the description of these kind of systems is still derived starting from a supergeometric setup and exploiting the so-called \emph{matter Ansatz}, constructed with a bosonic field and a spin-$1/2$ fermion field, invariant under Nieh-Yan-Weyl transformations \cite{Nieh:1981xk}. Models exhibiting Ususy do not require the matching of bosonic and fermionic degrees of freedom typical of supersymmetric theories; they involve dynamical spin-$1/2$ fermion fields and are particularly appealing because they are based on a Weyl-invariant action. Future investigations in this direction will be carried out under a field-theoretical perspective, at both classical and quantum level.

\section*{Acknowledgment}

We wish to thank Serena Fazzini for the inspiring discussions during the initial stages of this work. \\
D.M.P. acknowledges financial support
from the Chilean government through Fondecyt grants Grant $\#11240533$.

{
\normalsize 
 \bibliography{bosferm}
}
\end{document}